\documentclass[twocolumn,showpacs,prb,amsmath,amssymb,floatfix]{revtex4}

\usepackage{graphicx}
\usepackage{dcolumn}
\usepackage{bm}

\newcommand{\BFA}{BaFe$_2$As$_2$}
\newcommand{\BFP}{BaFe$_2$P$_2$}
\newcommand{\TCS}{ThCr$_2$Si$_2$}
\newcommand{\BFAP}{BaFe$_2$(As$_{1-x}$P$_x$)$_2$}
\newcommand{\BFAPx}{BaFe$_2$(As$_{0.71}$P$_{0.29}$)$_2$}
\newcommand{\BFAPy}{BaFe$_2$(As$_{0.53}$P$_{0.47}$)$_2$}
\newcommand{\BFAPz}{BaFe$_2$(As$_{0.24}$P$_{0.76}$)$_2$}
\newcommand{\BSFA}{(Ba$_{1-x}$Sr$_x$)Fe$_{2}$As$_{2}$}
\newcommand{\BSFAx}{Ba$_{0.5}$Sr$_{0.5}$Fe$_{2}$As$_{2}$}

\begin{document}



\title[\BFAP~and~\BSFA]{Different response of the crystal structure to isoelectronic doping in \BFAP~and \BSFA}

\author{Marianne Rotter, Christine Hieke}
\author{Dirk Johrendt}
\email{johrendt@lmu.de}\affiliation{Department Chemie, Ludwig-Maximilians-Universit\"{a}t M\"{u}nchen, Butenandtstrasse 5-13 (Haus D), 81377 M\"{u}nchen, Germany}

\date{\today}

\begin{abstract}

Superconductivity up to 30~K in charge neutrally doped \BFAP~has been ascribed to chemical pressure, caused by the shrinking unit cell. But the latter induces no superconductivity in \BSFA~in spite of the same volume range. We show that the spin-density-wave (SDW) state of \BFA~becomes suppressed in \BFAP~by a subtle reorganization of the crystal structure, where arsenic and phosphorus are located at different coordinates $z_{\rm{As}}$, $z_{\rm{P}}$. High-resolution X-ray diffraction experiments with \BFAP~single crystals reveal almost unchanged Fe--P bonds, but a contraction of the Fe--As bonds, which remain nearly unchanged in \BSFA. Since the Fe--As bond length is a gauge for the magnetic moment, our results show why the SDW is suppressed by P-, but not by Sr-doping. Only the Fe--P interaction increases the width of the iron 3$d$ bands, which destabilizes the magnetic SDW ground state. The simultaneous contraction of the Fe--As bonds is rather a consequence of the vanishing magnetism. Ordered structure models of \BFAP~obtained by DFT calculations agree perfectly with the single-crystal X-ray structure determinations. The contraction of the Fe--As bonds saturates at doping levels above $x \approx 0.3$, which corrects the unreasonable linear decrease of the so-called pnictide height.

\end{abstract}


\maketitle

\section{\label{intro}Introduction}

Superconductivity emerges in iron arsenides after the antiferromagnetic order of their parent compounds becomes suppressed \cite{Hosono-2008,Cruz-2008,Rotter-2008-1,Rotter-2008-2}. This proximity of magnetic and superconducting order parameters is widely accepted as one key argument for an unconventional pairing mechanism, which is mediated by spin fluctuations \cite{Cristianson-2008} as it was considered for the cuprates \cite{Shirane-1987,Keimer-1992}. But in strong contrast to the copper oxides, superconductivity in iron arsenides can be induced without changing the carrier concentration, either by applying external pressure \cite{Okada-2008,Alireza-2009} or by charge neutral doping.

Especially the isoelectronic substitution of arsenic by phosphorus in \textit{RE}OFe(As$_{1-x}$P$_x$) with \textit{RE} = La,~Ce \cite{CWang-2009, Cruz-2010, Luo-2010} and \textit{A}Fe$_2$(As$_{1-x}$P$_x$)$_2$ with \textit{A} = Ca,~Sr,~Ba,~Eu \cite{Ren-2009,Shi-2009,Jiang-2009} has turned out to be an ideal tuning parameter to explore the borderline of magnetic ordering and superconductivity. In iron arsenides, the already present Fermi surface can be finely tuned by the crystal structure parameters in order to allocate optimal conditions for superconductivity. However, the detailed interplay between the crystal structure, magnetic ordering and superconductivity is hardly understood up to now, which is to some extent due to the lack of precise structural data.

One of the most intensively investigated iron-based material is \BFA~with the tetragonal \TCS-type structure \cite{Rotter-2008-1}. The stoichiometric compound undergoes a structural phase transition, associated with antiferromagnetic ordering (spin-density-wave, SDW) at $T_{\rm{tr}}$ = 140~K. Superconductivity emerges during the antiferromagnetic ordering gets suppressed, which can be achieved by  hole- or electron-doping \cite{Rotter-2008-2, Sefat-2008} or by applying external pressure \cite{Alireza-2009}.

From the latter it seems easy to conceive, that also the so-called 'chemical pressure' by substituting arsenic for the smaller phosphorus atoms induces superconductivity in LaOFe(As$_{1-x}$P$_x$)\cite{CWwang-2009} and \BFAP~\cite{Jiang-2009}. The authors explain the appearance of superconductivity as a consequence of the shrinking unit cell volume in analogy to the effects of external pressure. On the other hand, the unit cell shrinks comparably by substituting barium for smaller strontium atoms in \BSFA, but no superconductivity appears in this case \cite{Wang-2009}. Thus, a pressure-volume effect is clearly an oversimplified explanation for superconductivity in the case of phosphorus-doped \BFAP. This is the more true, as the crystal structures of these compounds have not yet been investigated in detail. We have therefore synthesized the series \BFAP~and \BSFA~and determined the crystal structures by high-resolution X-ray powder and single crystal diffraction. We compare the structural data with theoretical models obtained by full-potential DFT calculations. We will show that the suppression of the SDW ordering as precondition to the onset of superconductivity is not a simple volume effect, but depends on a subtle reorganization of the crystal structure.

\section{Methods}

\begin{figure}
\includegraphics[width=70mm]{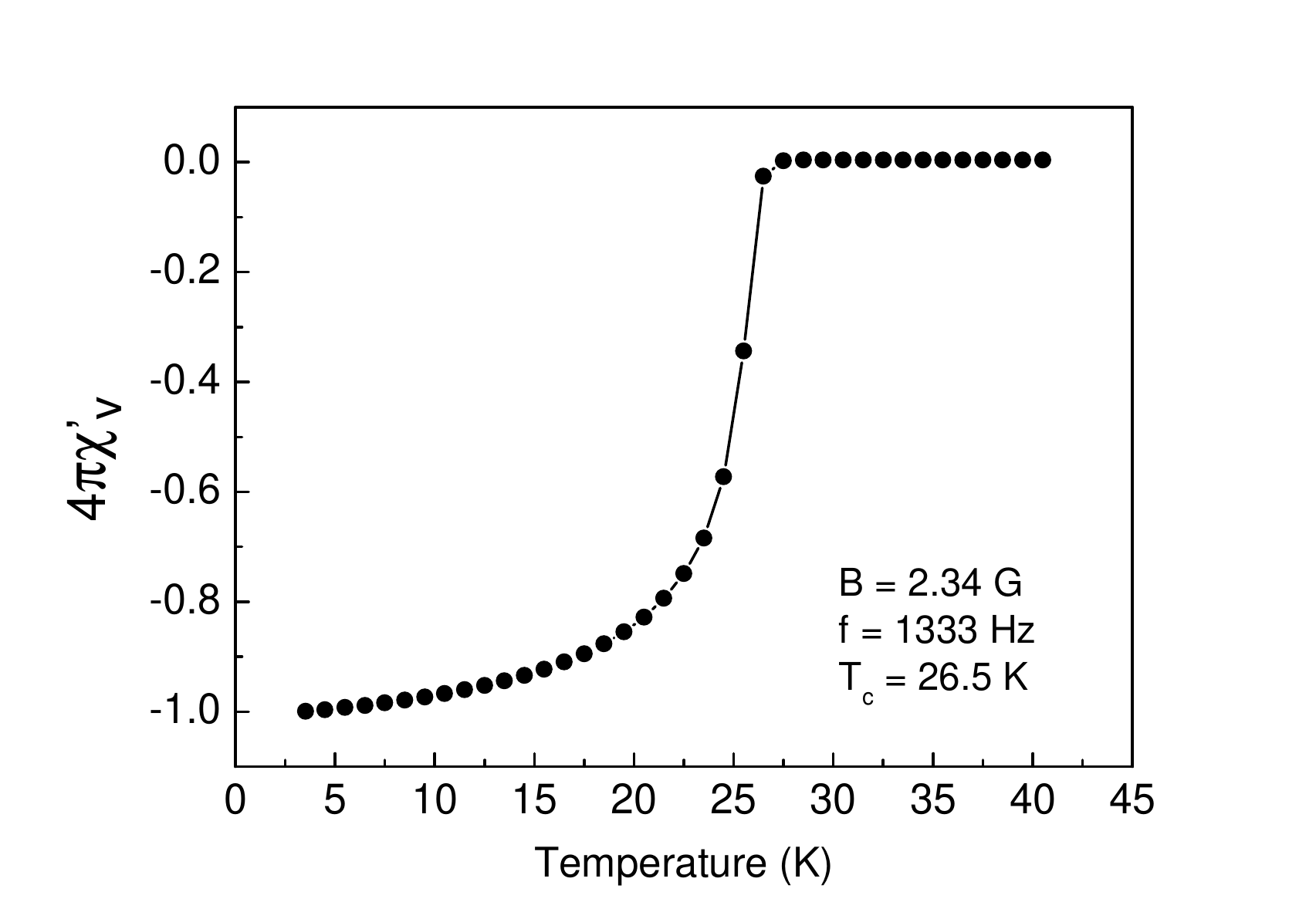}
\caption{\label{fig:suszept} AC-susceptibilty of \BFAP~ with $x \approx 0.4$}
\end{figure}

Polycrystalline samples of \BSFA~($x$ = 0-1) were synthesized by heating stoichiometric mixtures of the elements (all purities $>$ 99.9\%) in alumina crucibles enclosed in silica tubes under an atmosphere of purified argon. The mixtures were slowly heated to 1123~K, kept at this temperature for 15~h and cooled down to room temperature. The reaction products were homogenized in an agate mortar and annealed three to four times at 1173~K for 15~-~25~h. The obtained black metallic powders are slightly air sensitive and therefore handled under argon atmosphere. Samples of \BFAP~with $x$ = 0-1 were synthesized by solid-state reaction of the elements likewise. The stoichiometric mixtures were slowly heated to 1123~K (1173 K) for 15~h. After cooling down to room temperature, the products were ground and annealed at 1173~K for 25~h (two to three times at 1273~K and 1323~K, respectively). The obtained black crystalline powders show no sensitivity to air or moisture. EDX measurements resulted in the nominal compositions within 10\% regarding to the Ba:Sr and As:P ratios. Samples between $x \approx$ 0.3 and 0.6 are superconducting with critical temperatures up to 29~K. No superconductivity was found in \BSFA. Since both phase diagrams are known \cite{Jiang-2009,Wang-2009}, we show only the AC-susceptibility measurement of a \BFAP~sample with $x \approx 0.4$ as an example in Figure~\ref{fig:suszept}.

The crystal structures of polycrystalline samples were determined by X-ray powder diffraction using a Huber G670 Guinier imaging plate diffractometer (Cu-K$\alpha_1$ radiation, Ge-(111) monochromator). Rietveld refinements were performed with the TOPAS package \cite{Topas} using the fundamental parameters approach as reflection profiles. Due to a better description of small peak half width and shape anisotropy, the approach of \textit{Le Bail} and \textit{Jouanneaux} \cite{LeBail} was implemented in the TOPAS program. Giving consideration to possible texture effects, the March-Dollase function was used for description of a preferred orientation of the crystallites.

Small plate-like single crystals of $\approx 50 \times 50 \times 20$ microns were selected from the polycrystalline samples and checked by Laue photographs using white radiation from a Mo-anode. Diffraction intensity data up to $2\theta$ = 80$^{\circ}$ were collected with an Oxford Xcalibur 4-circle $\kappa$-diffractometer equipped  with a CCD detector. Graphite-monochromized Mo-K$\alpha$ radiation from a conventional sealed tube was used. The intensities were carefully corrected for absorption effects. The atom positions from \BFA~\cite{Rotter-2008-1} were used as starting parameters and refined by the least squares method using the Shelxl program package \cite{Shelx}. Positional and isotropic displacement parameters of As and P were refined independently, while their occupation parameters were constraint to unity.


Electronic structure calculations were performed with the WIEN2k package \cite{Blaha-2001} using density functional theory within the full-potential LAPW method and the general gradient approximation (GGA). Detailed descriptions are given elsewhere \cite{Schwarz-2003}. LAPW is based on the muffin-tin construction with non-overlapping spheres and a plane wave expansion in the interstitial regions. Mixed LAPW and APW~+~lo (lo = local orbitals) basis sets were used to increase the efficiency of the APW linearization \cite{Sjostedt-2000}. Further technical details are given in the monograph of \textit{Singh} \cite{singh-2006}.

\section{Results and Discussion}

\subsection{SDW suppression in \BFAP~and \BSFA}

\begin{figure}
\includegraphics[width=65mm]{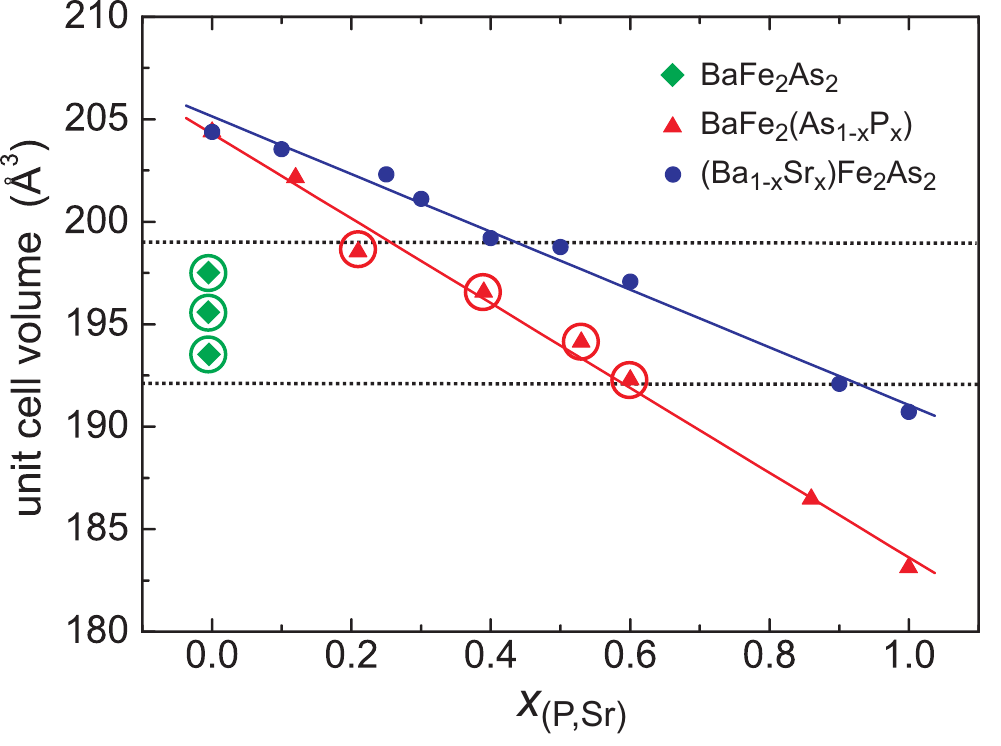}
\caption{\label{fig:Volumenplot} Unit cell volume against the doping levels in \BFAP~and \BSFA. Superconducting samples are marked by circles. Values of undoped \BFA~under pressure are from ref.~\cite{Kimber-2009} (Color online).}
\end{figure}

\begin{figure}
\includegraphics[width=70mm]{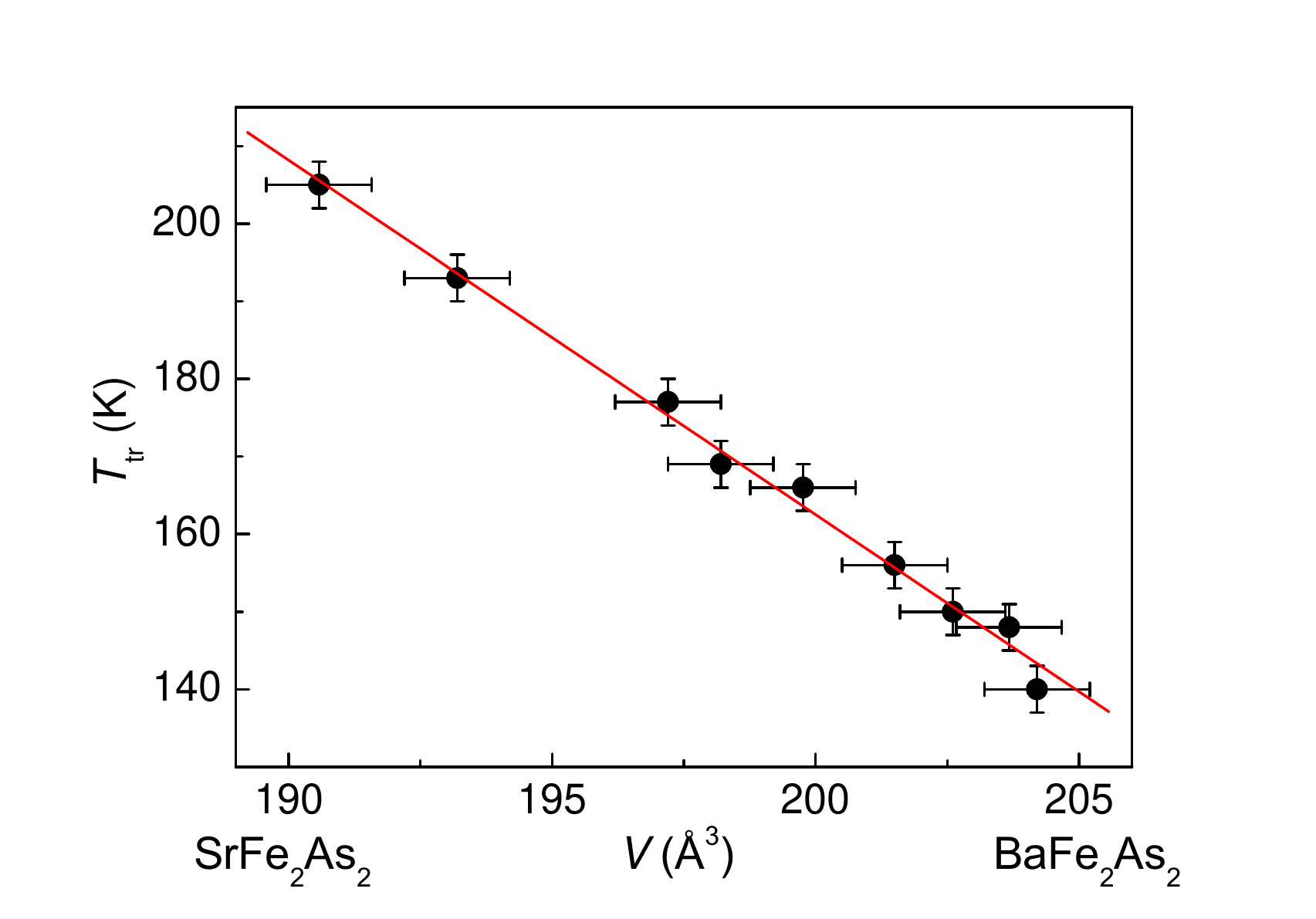}
\caption{\label{fig:Ttr_vol} SDW-transition temperatures of \BSFA~ (Color online).}
\end{figure}

The unit cell volumes of the solid solution series \BFAP~and \BSFA~together with the data of the high-pressure experiments by \textit{Kimber} et al. \cite{Kimber-2009} are collected in Figure~\ref{fig:Volumenplot}. Superconductivity (indicated by circles) indeed appears by pressure and by P-doping within the same volume range between 192 and 199~{\AA}$^3$, which is marked by the dashed lines. But on the other hand, no superconductivity emerges by Sr-doping in \BSFA, even though the unit cell volumes are within the same range. Obviously, the simple shrinking of the unit cell is not the decisive condition. Superconductivity is absent in \BSFA, because Sr-doping cannot suppress the magnetic and structural transition. The latter becomes only shifted to higher temperatures, as depicted in Figure~\ref{fig:Ttr_vol}.

\begin{figure}
\includegraphics[width=70mm]{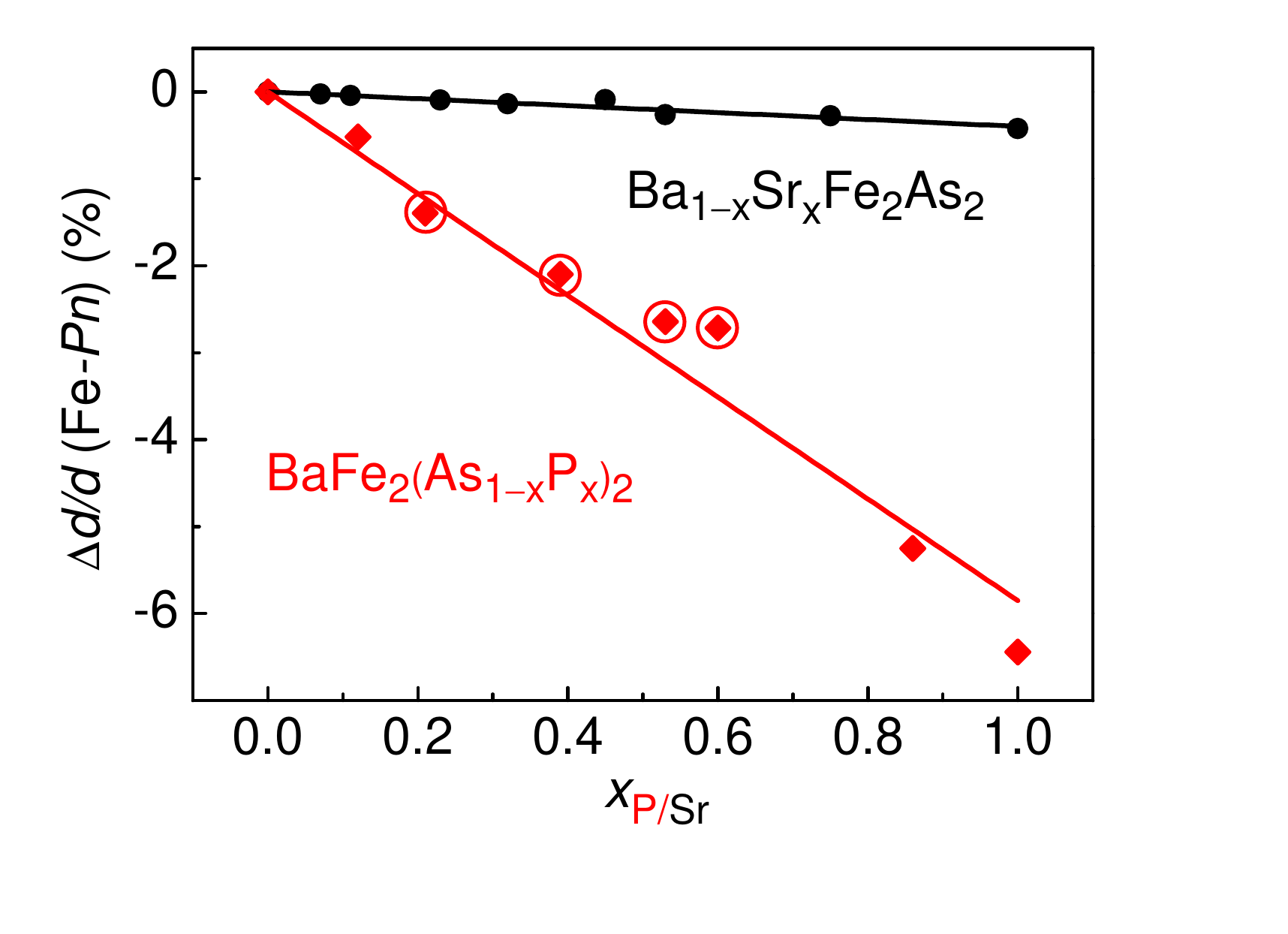}
\caption{\label{fig:DeltaD-x} Normalized iron to pnictogen ($Pn$ = P,~As) distances of \BSFA~and \BFAP, obtained from Rietveld-refinements. Superconducting samples are marked by circles (Color online).}
\end{figure}

In order to shed light on this apparent contradiction, a detailed look at the crystal structures is indispensable. First, we have refined the X-ray powder data by the Rietveld-method. The pnictogen ($Pn$ = As,~P) atoms occupy the 4e (00$z$)-position, where $z_{\rm{Pn}}$ is the only variable parameter of the tetragonal \TCS-type structure (space group $I4/mmm$). The doping dependencies of the normalized Fe--$Pn$ bond lengths are plotted in Figure~\ref{fig:DeltaD-x}. Sr- and P-doped \BFA~show very different behavior, respectively. While the Fe--As bond lengths in \BSFA~remain almost constant across the whole doping range despite the shrinking unit cell, the Fe--$Pn$ distances decrease strongly with increasing P-content in \BFAP. This strongly suggests that a shortening of the Fe--$Pn$ bonds is necessary to suppress the SDW transition.

\begin{figure}
\includegraphics[width=85mm]{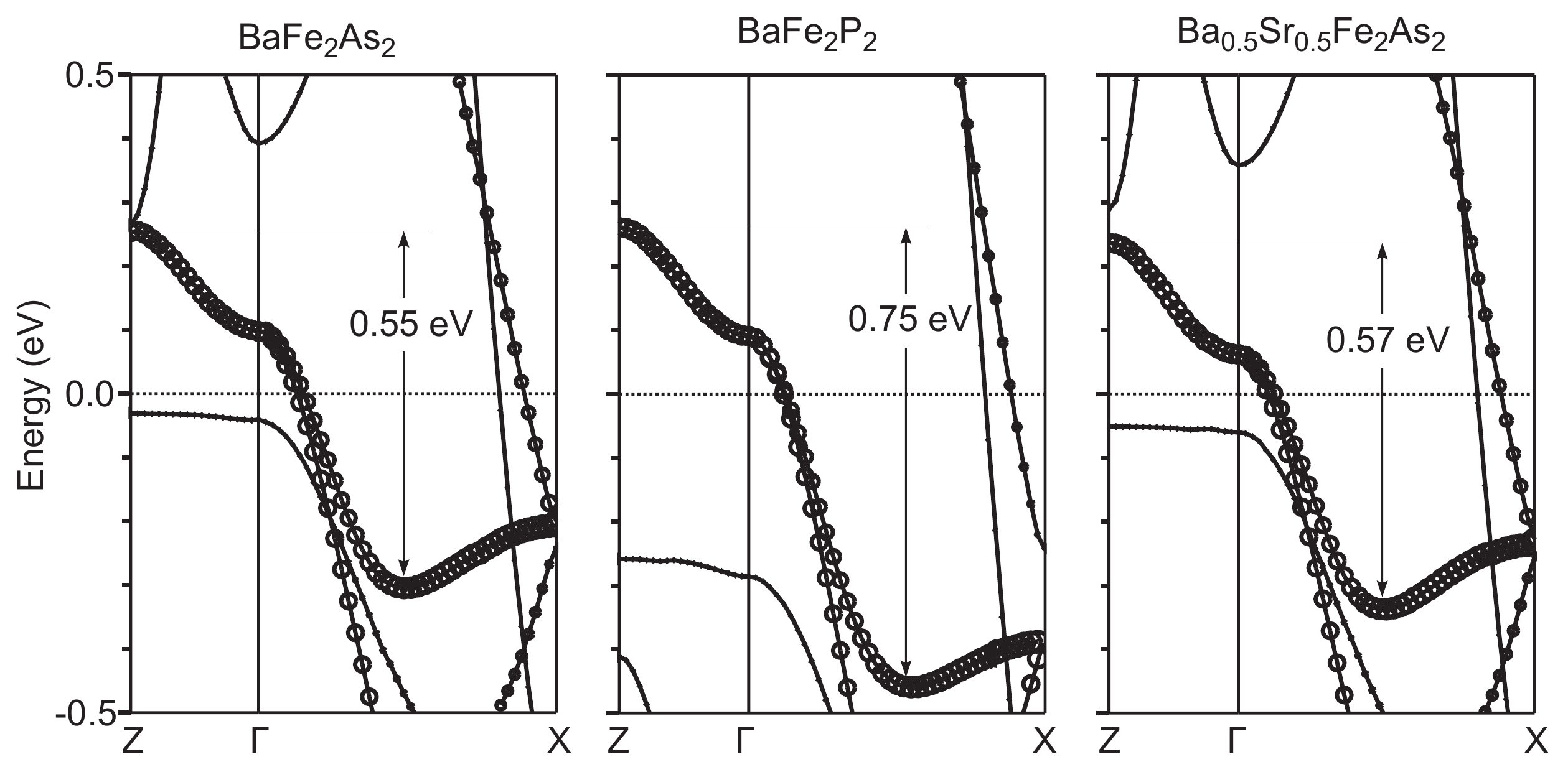}
\caption{\label{fig:bands} Band structures of \BFA, \BFP~and \BSFAx. The bandwidths of the Fe-$d_{xz,yz}$ bands are indicated.}
\end{figure}

The latter is plausible, since it is well known that the Fe--As bond length dramatically influences the magnetism in iron arsenides \cite{Egami-2010, Johannes-2010}. As a first approach, we assume that decreasing bond lengths increase the bandwidth, thus a magnetic ground-state becomes less stable as the bonds get shorter. Figure \ref{fig:bands} shows sections of the band structures with the approximate bandwidths of the Fe-$d_{xz,yz}$ bands, which should be strongly effected by the Fe--$Pn$ interaction. We find that the bandwidth of \BFP~is much larger (+35\%) compared with \BFA. But in the case of \BSFAx, the bandwidth is almost identical to \BFA, even though the volume is significantly smaller.

\subsection{Dependency of the magnetic moment on the Fe--As distance}

\begin{figure}
\includegraphics[width=70mm]{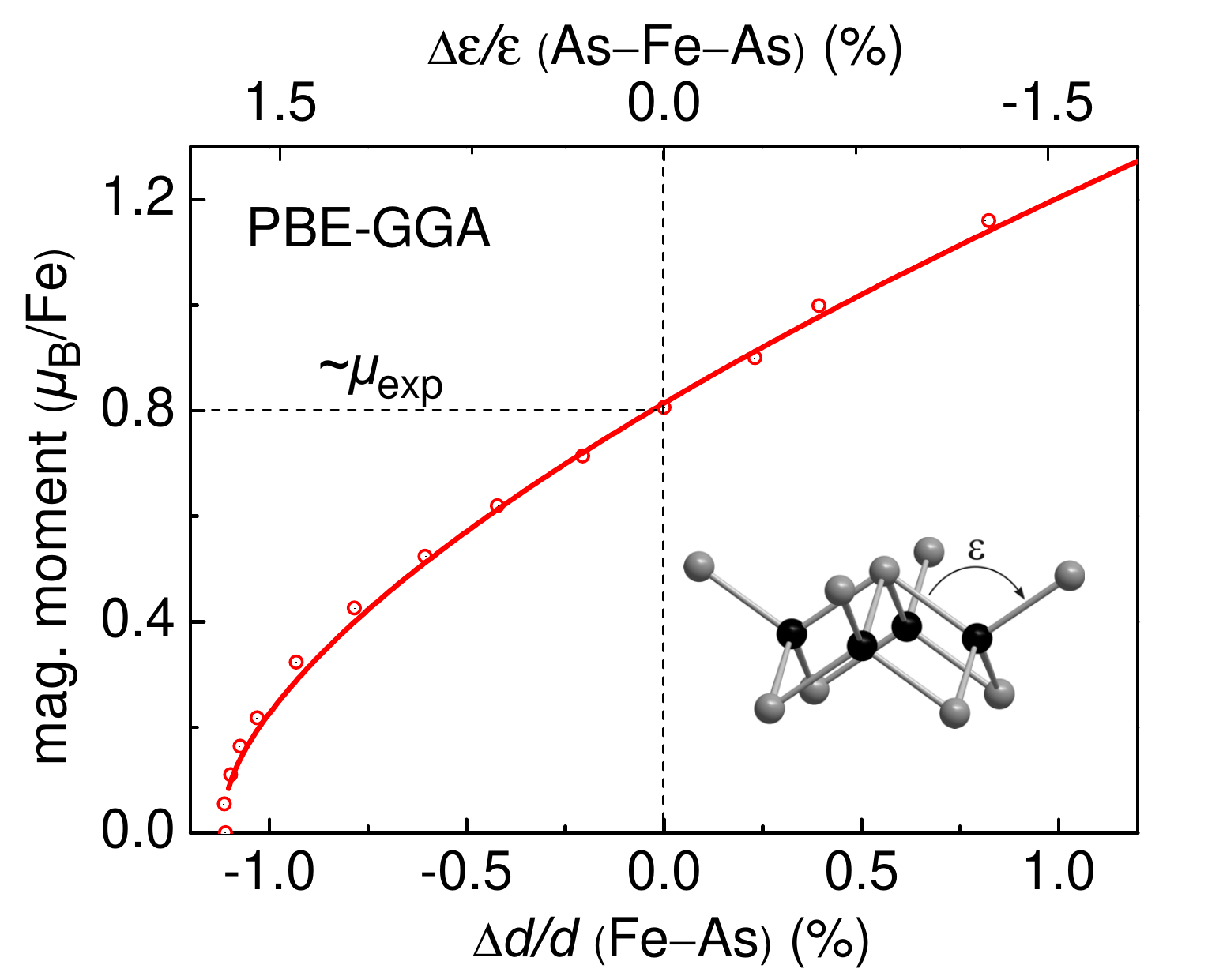}
\caption{\label{fig:mu-DeltaD}Magnetic moments by variation of the Fe--As bond length in \BFA~(Color online).}
\end{figure}

To support this assumption, we have calculated the dependency of the magnetic moment on the Fe--As bond lengths using a full-potential DFT method. It is well known that DFT is unable to describe the exact value of the magnetic moment and the correct structure at the same time \cite{Mazin-2008-2}. But on the other hand, it should be possible to get at least the changes of the magnetic moment as the Fe--As distance decreases. Thus we have selected a $z_{\rm{As}}$ value, which gives approximately the experimental moment of 0.8~$\mu_{\rm{B}}$/Fe. Then we used the fixed-spin-moment method to calculate $z$-values by varying the magnetic moments between 0 and 1.2~$\mu_{\rm{B}}$/Fe and relaxing the structures subsequently. We find that the moment gets rapidly depleted as the Fe--As distance decreases (see Figure~\ref{fig:mu-DeltaD}) and a shrinking of 2.5~pm ($\approx 1\%$) is sufficient to suppress the magnetism completely. This is a rather simplified approach, since we used a ferromagnetic spin structure and kept the volume constant. For this reason, also the As--Fe--As angle varies about 1.5\%, thus it is not a priori clear whether the bond angle or bond distance is the crucial parameter.

\begin{figure}
\includegraphics[width=85mm]{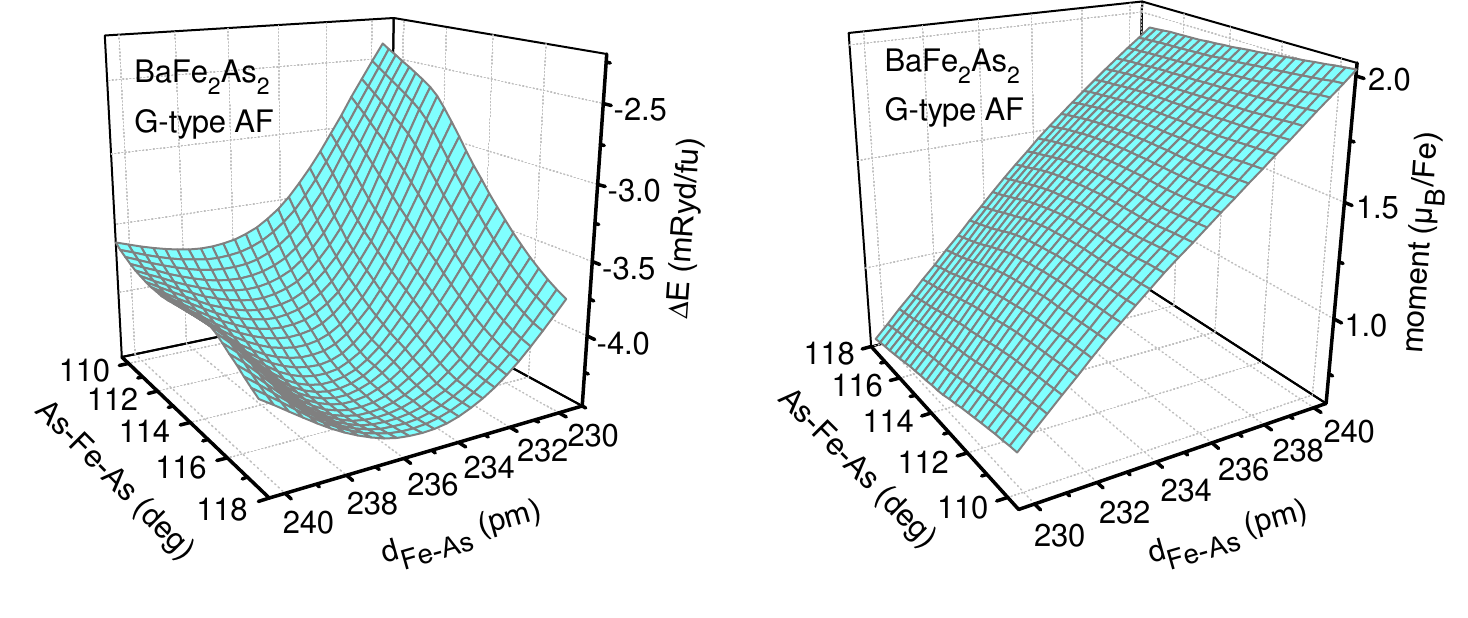}
\caption{\label{fig:BFA_AF_OPT} Results of the full structural optimization of \BFA. The antiferromagnetism is approximated by a simple $G$-type spin structure (Color online).}
\end{figure}

To check the validity of our simple model, we performed full optimizations of the structure (volume, $z$, $\mu$) using a $G$-type spin structure in order to model the AF magnetic interactions in the tetragonal phase. Figure~\ref{fig:BFA_AF_OPT} shows how the total energy and the magnetic moment depend on the bond distances and angles. The energy surface has a minimum somewhat beyond the experimental data (as usual for \BFA), here also owing to the crude approximation of the magnetic structure. More important is the fact, that the magnetic moment varies strongly with the Fe--As distance, but only slightly with the As--Fe--As angle. This model, although still simplified, indicates that the Fe--As distance is the crucial parameter which determines the magnetic moment, in agreement with recent results reported by \textit{Johannes} and \textit{Mazin} \cite{Johannes-2010}. Thus, we can assume that the small shifts of the distance plotted in Figure~\ref{fig:mu-DeltaD} are associated with the suppression of the SDW state.

It is worthwhile to note, that the suppression of the SDW in \BFAP~may not necessarily be \textit{caused} by the shrinking Fe--As bonds. One may also argue that the increasing number of Fe--P contacts enlarges the bandwidth, until the magnetic ground-state becomes unstable. In this case, the contraction of the Fe--As bonds would be the consequence of the lost magnetism rather than its origin. However, this interpretation emphasizes that the Fe--As bond is in any case the gauge for the magnetic moment, as pointed out recently \cite{Johannes-2010}.

By comparing these results with the data in Figure~\ref{fig:DeltaD-x}, we find that the shrinking of the Fe--As distances in \BSFA~is only $\approx$~1~pm (-0.4\%) and thus not sufficient to suppress the magnetism according to Figure~\ref{fig:mu-DeltaD}. On the other hand, the decrease of the Fe--$Pn$ bond lengths is much larger in \BFAP~and evidently sufficient to suppress the SDW. Moreover, the calculated value of $\frac{\Delta d}{d} \approx -1~\%$ coincides perfectly with the onset of superconductivity in \BFAP~close to the doping value $x \approx 0.25$ as shown in Figure~\ref{fig:DeltaD-x}.

Even though this scenario explains the difference between the Sr- and P-doped materials with respect to the onset of superconductivity, the data of \BFAP~in Figure~\ref{fig:DeltaD-x} have no physical meaning. Phosphorous and arsenic cannot have the same coordinates in the unit cell, otherwise the atomic distances would be unreasonable. As an example, the data of Figure~\ref{fig:DeltaD-x} anticipate that the Fe--As \textit{and} Fe--P bonds at $x \approx 0.25$ are about 236~pm. This is still reasonable for a Fe--As bond, but much too long for a Fe--P bond. On the other hand, both distances would be 228~pm at $x \approx 0.75$, which is now acceptable for Fe--P, but impossible for an Fe--As bond.

\subsection{Theoretical structure models}

In order to to investigate the structural behavior of \BFAP~theoretically, we used crystallographically ordered super-structures. The tetragonal space group $P\overline{4}m2$ allows to calculate structural parameters for the doping levels $x$ = 0.25, 0.5 and 0.75 without changing the symmetry. During the optimization of the atom coordinates, all compositions except $x$ = 0 converged to non-magnetic ground-states. Thus, the SDW state is already suppressed at $x$ = 0.25 in agreement with the experiment. Furthermore, the $z_{\rm{Pn}}$-coordinates of arsenic and phosphorus become significantly different, which lead to longer Fe--As and shorter Fe--P bonds, as expected from atom size considerations. Selected bond distances are collected in Table~\ref{tab:calculated}.

\begin{table}[h]
\caption{\label{tab:calculated} Bond lengths of \BFAP~from DFT calculations}
\begin{ruledtabular}
\begin{tabular}{llllll}
$x$ = & 0 & 0.25 & 0.50 & 0.75 & 1 \\

\hline
Fe~--~As & 235.9 & 232.7 & 232.7 & 232.6   & -     \\
Fe~--~P  &  -    & 222.0 & 222.5 & 221.1   & 221.2 \\
Ba~--~As & 346.0 & 341.2 & 337.2 & 335.7   &  -    \\
Ba~--~P  &  -    & 354.5 & 349.3 & 342.9   & 338.0 \\
\end{tabular}
\end{ruledtabular}
\end{table}

It is evident that the Fe--As bonds shrink significantly only between $x$ = 0 and 0.25, accompanied by the loss of the magnetic moment. The contraction is about {-1.4\%}, which is sufficient to suppress the SDW order as shown in Figure~\ref{fig:mu-DeltaD}. But higher doping levels do not further contract the Fe--As bonds, in contrast to the data in Figure~\ref{fig:DeltaD-x}. At the same time, the Fe--P bonds remain close to the values of the pure phosphide. These findings draw a more realistic picture than Figure~\ref{fig:DeltaD-x}, because it combines the suppression of the SDW state by the appropriate contraction of the Fe--As bond with reasonable bond lengths along the whole doping range.

\subsection{Single crystal structure determination}

\begin{figure}
\includegraphics[width=80mm]{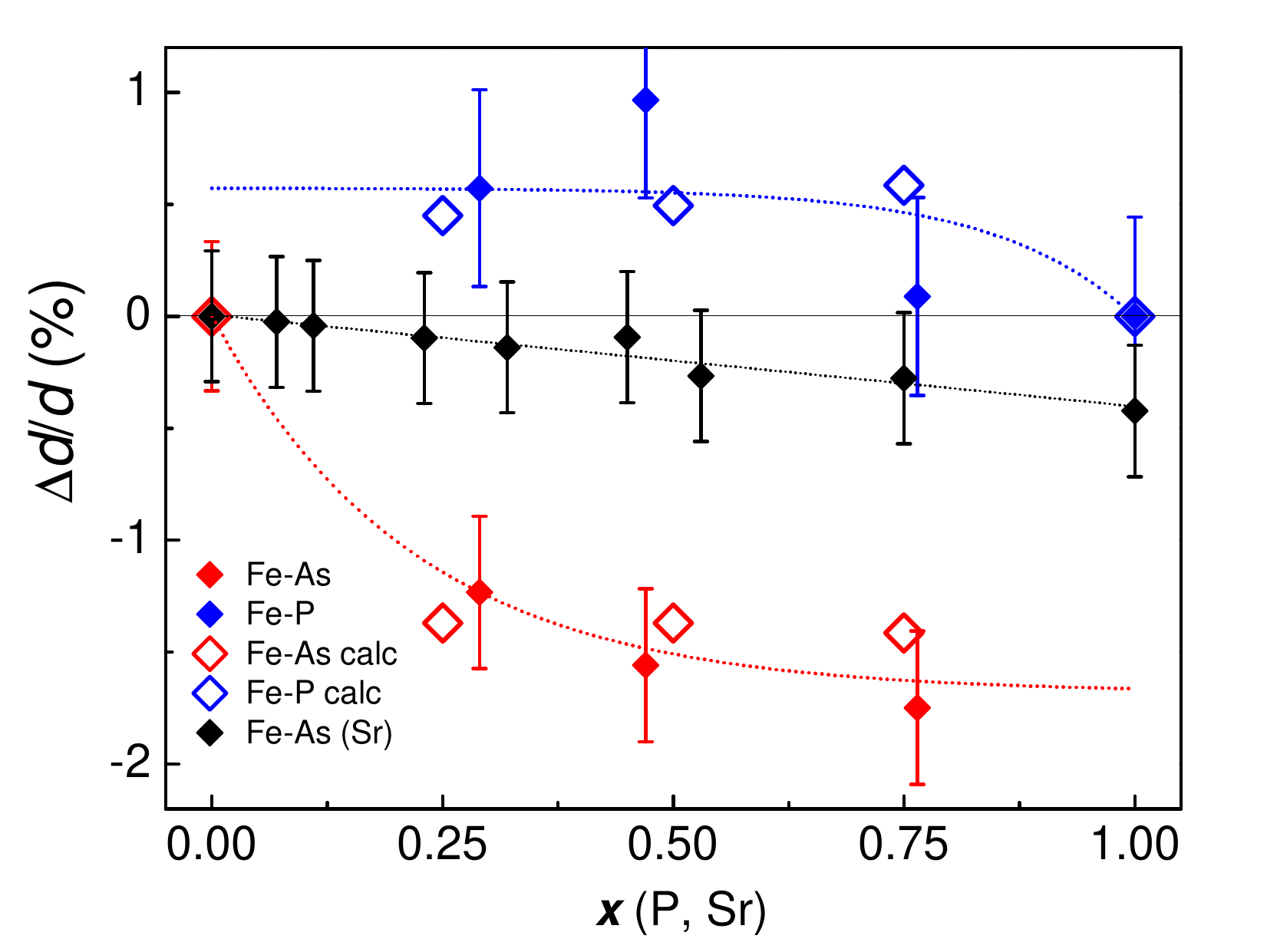}
\caption{\label{fig:bonds} Changes of the normalized Fe--As and Fe--P bond lengths in \BFAP~from single crystal data and in \BSFA~from powder diffraction. Error bars represent four times the estimated standard deviations. Open symbols represent Fe--$Pn$ bond lengths predicted by DFT structure optimizations of \BFAP. Dotted lines are guides to the eye.}
\end{figure}

From the crystal chemical considerations given in section B. and the results of the theoretical calculations it is clear that the Fe--\textit{Pn} bond lengths of \BFAP~obtained from Rietveld refinements are throughout unsatisfactory, whereas the Fe--As distances in \BSFA~are reasonable due to only one arsenic position. For this reason, we have conducted high-resolution single crystal X-ray structure determinations using small crystals (tenth of microns) of \BFAP~with $x~\approx$~0.3, 0.5 and 0.7. Main results of the refinements together with selected bond lengths and angles are summarized in Table~\ref{tab:Crystallographic}. As expected from chemical reasons as well as predicted by theory, we find the P- and As-atoms at significantly different $z$-coordinates, resulting in different lengths of the Fe--As and Fe--P bonds. The initial Fe--As distance is 240~pm and decreases rapidly up to $\approx$~25\% P-doping, but then converges to an almost constant value around 236~pm (-1.7\%). At the same time, the Fe--P distances remain about 1-2~pm longer than in \BFP~at least up to 75\% P-doping. These experimental findings agree almost quantitatively with the theoretical prediction, although theory is not able to produce the correct absolute values. Figure~\ref{fig:bonds} summarizes this reorganization of the crystal structures of \BFAP~in comparison with \BSFA. A small contraction of the Fe--As bonds by $\approx$~-1.5\% (2-3~pm) is associated with the suppression of the SDW state, which is in turn the precondition to superconductivity. This happens by P-, but not by Sr-doping, where the Fe--As bonds shrink by less than 0.5\%.

Even though we have focused at the Fe--As bonds, we suggest that the Fe--P interaction is the underlying origin of the SDW suppression. Thus we have rather detected the response of the structure to the vanishing magnetic state. Interestingly, a similar response was not observed in undoped \BFA~under pressure \cite{Kimber-2009}, where superconductivity emerges despite the Fe--As bonds remain constant as in non-superconducting \BSFA.

\section{Conclusion}

We have shown that 'chemical pressure' is an oversimplified view, which is unsuitable to explain the different behavior of charge neutrally doped \BFAP~and \BSFA. The true reason is the different response of the crystal structures to doping. Phosphorus doping suppresses the SDW state by increasing the width of the $d$-bands, which in turn leads to shorter Fe--As bonds due to its strong coupling to the magnetic state. No significant contraction of the Fe--As bond is detected in \BSFA, where the SDW state persists despite the similar volume change. Single crystal X-ray data revealed the correct structure of \BFAP, where arsenic and phosphorus are statistically distributed, but at different coordinates $z_{\rm{As}}$ and $z_{\rm{P}}$, which is necessary to obtain reasonable Fe--P and Fe--As bond lengths. The latter contract strongly at low doping during the magnetism is suppressed ($0 \leq x \leq 0.25$) and converges to $\approx$~236~pm at higher doping levels. This behavior is almost quantitatively reproduced by DFT calculations using ordered model structures of \BFAP. Our results strongly emphasize, that even subtle details of the crystal structures are crucial to magnetism and superconductivity in iron-based materials, where structural data have often been obtained from moderate powder patterns, which allow only limited resolution.

\begin{acknowledgments}

We thank Franz A.~Martin for measurements with, and Prof. Thomas Klap\"{o}tke for access to the single crystal diffractometer. Marcus Tegel is acknowledged for the AC-susceptibility measurements. This work was supported financially by the DFG (German Research Foundation).

\end{acknowledgments}

\begin{table*}[p]
\caption{\label{tab:Crystallographic} Crystallographic data of \BFAP~(Space group $I4/mmm$, $Z$ = 2, Mo-K$_{\alpha}$ radiation, $\lambda$ = 71.073~pm)}
\begin{ruledtabular}
\begin{tabular}{llllll}
 Compound                & \BFA           & \BFAPx        & \BFAPy        & \BFAPz       & \BFP     \\
 Molec. mass (g/mol)~    & 398.88         & 372.95        & 357.63        & 331.86       & 319.77    \\
 \textit{a} (pm)         & 396.24(1)      & 391.78(1)     & 390.65(1)     & 386.60(3)    & 384.35(4) \\
 \textit{c} (pm)         & 1301.35(8)     & 1276.10(7)    & 1273.55(6)    & 1259.2(1)    & 1242.2(2) \\
 \textit{V} (nm$^{3}$)   & 0.2043(1)      & 0.1959(1)     & 0.1944(2)     & 0.1892(4)    & 0.1835(2) \\
 Abs. coeff. (mm$^{-1}$) & 32.4           & 29.1          & 27.1          & 22.5         & 20.7      \\
 $R_{int}, R_{\sigma}$   & 0.041, 0.024   & 0.030, 0.013  & 0.035, 0.016  & 0.050, 0.021 & 0.093, 0.042 \\
 Ref. param., GooF       & 9, 1.09        & 10, 1.19      & 10, 1.06      & 10, 1.15     & 8, 0.997 \\
 $R1, wR2$ (all data)  & 0.040, 0.070~~   & 0.023, 0.049  & 0.020, 0.035  & 0.033, 0.082 & 0.031, 0.060 \\
 As $z$ position         & 0.35385(7)     & 0.3544(1)     & 0.3542(1)     & 0.3571(4)    & -\\
 P  $z$ position         & -              & 0.3402(1)     & 0.3438(4)     & 0.3430(4)    & 0.3459(1) \\
Bond lengths (pm):\\
Ba--As                   & 338.64(6)     & 333.58(7)      & 332.2(1)      & 327.3(2)     & -         \\
Ba--P                    & -             & 344.0(2)       & 339.6(3)      & 337.4(2)     & 332.4(1)  \\
Fe--As                   & 239.82(6)     & 236.89(7)      & 236.1(1)      & 235.7(3)     & -         \\
Fe--P                    & -             & 227.2(4)       & 228.9(3)      & 226.0(2)     & 226.14(9) \\
Bond angles (deg):\\
As--Fe--As               & 111.40(3)     & 111.57(5)      & 111.6(2)      & 110.2(2)     & -         \\
As--Fe--P                & -             & 115.4(3)       & 114.4(1)      & 113.9(1)     & -         \\
P--Fe--P                 & -             & 119.1(3)       & 117.1(2)      & 117.6(1)     & 116.39(7) \\
\end{tabular}
\end{ruledtabular}
\end{table*}

\bibliography{BFAP}

\end{document}